\documentclass[onecolumn]{emulateapj}
\usepackage{graphicx}
\usepackage{amsfonts,amsmath,amssymb,mathrsfs}
\usepackage{color}

\newcommand{\be}{\begin{eqnarray}}
\newcommand{\ee}{\end{eqnarray}}

\begin{document}

\title{Constraints on the spacetime geometry around 10~stellar-mass black hole candidates from the disk's thermal spectrum}

\author{Lingyao Kong, Zilong Li, and Cosimo Bambi\footnote{Corresponding author: bambi@fudan.edu.cn}}

\affil{Center for Field Theory and Particle Physics \& Department of Physics, Fudan University, 200433 Shanghai, China}

\begin{abstract}
In a previous paper, one of us has described a code to compute the thermal 
spectrum of geometrically thin and optically thick accretion disks around
generic stationary and axisymmetric black holes, which are not necessarily 
of the Kerr type. As the structure of the accretion disk and the propagation of 
electromagnetic radiation from the disk to the distant observer depend on 
the background metric, the analysis of the thermal spectrum of thin disks 
can be used to test the actual nature of black hole candidates.
In this paper, we consider the 10 stellar-mass black hole candidates for which 
the spin parameter has been already estimated from the analysis of the disk's 
thermal spectrum and under the assumption of the Kerr background, and we 
translate the measurements reported in the literature into 
constraints on the spin parameter--deformation parameter
plane. The analysis of the disk's thermal 
spectrum can be used to estimate only one parameter of the geometry close 
to the compact object, and therefore it is not possible to get independent 
measurements of both the spin and the deformation parameters. 
The constraints obtained here will be used in 
combination with other measurements in future work, with the final goal to 
break the degeneracy between the spin and possible deviations from the Kerr 
solution and thus test the Kerr black hole hypothesis. 
\end{abstract}



\section{Introduction}

General relativity makes very clear predictions about the properties of the 
spacetime geometry around a black hole (BH). According to the no-hair 
theorem~\citep{nh1,nh2,nh3}, 4-dimensional uncharged BHs are only 
described by the Kerr solution, which is completely specified by two parameters, 
associated respectively to the mass $M$ and the spin angular momentum 
$J$ of the compact object. $a_* = J/M^2$ is the spin parameter and a Kerr
BH must have $|a_*| \le 1$. For $|a_*| > 1$, there is no event horizon and there 
is a number of theoretical arguments suggesting that such super-spinning Kerr 
objects can unlikely be relevant in astrophysics~\citep{ss1,ss2,ss3,ss4}.

Astronomical observations have discovered at least two classes of BH candidates.
Dark and compact objects in X-ray binary systems with a mass $M \approx 5 - 
20$~$M_\odot$ \citep{bh1} are too heavy to be relativistic stars made of neutrons
or quarks for plausible matter equations of state~\citep{bh3,bh3b}. The supermassive 
BH candidates at the center of every normal galaxy have a mass $M \sim 10^5 - 
10^9$~$M_\odot$ \citep{bh2} and they turn out to be too massive, compact, and 
old to be clusters of non-luminous bodies, because the cluster lifetime due to physical 
collisions and evaporation would be shorter than the age of these systems~\citep{bh4}.
For both stellar-mass and supermassive BH candidates, there is no evidence 
of electromagnetic radiation emitted by their surface, which may be interpreted 
as an indication for the presence of an event/apparent horizon~\citep{eh0,eh1,eh2,eh3}.
In the end, all these objects are thought to be the Kerr BHs of general relativity 
simply because they cannot be explained otherwise without introducing new physics.

Despite this body of indirect evidence, there is no indication that the spacetime 
geometry around BH candidates is described by the Kerr metric. The Kerr BH
hypothesis entirely relies on the validity of standard physics. The no-hair theorem 
might be avoided by a number of ways, such as by considering exotic forms of 
matter, non-stationary solutions, or non-trivial extensions of general relativity. It 
is therefore of extreme importance to experimentally test the Kerr nature of 
astrophysical BH candidates and, thanks to the progresses in the last decade in 
the understanding of the accretion processes and in the observational facilities, 
such a possibility is not out of reach any more [for a review, see e.g.~\citet{rev1,rev2}].
This goal can be achieved by extending the techniques that have been developed 
to estimate the spin parameter of BH candidates under the assumption that they 
are of the Kerr type. The compact object is now specified by a mass $M$, a 
spin parameter $a_*$, and at least one deformation parameter, which measures 
possible deviations from the Kerr geometry. One then computes the observational
features in this more general background. The comparison of the theoretical 
predictions with the observational data is now used to estimate both the spin 
and the deformation parameters. If one finds a vanishing deformation parameter,
the Kerr BH hypothesis is verified. If the observational data point out a non-vanishing
deformation parameter, the Kerr nature of astrophysical BH candidates is 
questioned. However, even assuming to have the correct astrophysical
model and all the systematic effects under control, there is usually
a strong correlation between the estimate of the spin and of the deformation 
parameter, so the final result is a constraint on the spin parameter--deformation 
parameter plane~\citep{cfm1,v2-k,iron-cb,v2-cb}. 
Only in presence of excellent data, not available today and probably 
not even in the near future, it might be possible to solve such a fundamental 
degeneracy between the spin and possible deviations from the Kerr geometry.
The strategy is therefore to combine different measurements of the same object and
break this degeneracy~\citep{shadow2,1409}. 
With a single measurement, we can only rule out some BH
alternatives with very specific features, like some exotic compact objects without 
event horizon~\citep{boson1,boson2} and some wormholes~\citep{wh}, or 
constrain the deformation parameter in classes of non-Kerr BHs that can mimic
fast-rotating Kerr BHs only for a restricted range of the spin and of the 
deformation parameters~\citep{bardeen}.

At present, the most robust technique to probe the spacetime geometry around 
stellar-mass BH candidates is probably the continuum-fitting method; that is, the
analysis of the thermal spectrum of geometrically thin and optically thick accretion 
disks~\citep{zhang}. In the case of Kerr BHs, the thermal spectrum of thin 
accretion disks depends on 5 free parameters; that is, the BH mass $M$, the BH 
spin parameter $a_*$, the mass accretion rate $\dot{M}$, the BH distance $d$, 
and the inclination angle $i$ of the disk with respect to the line of sight of the
distant observer. As put forward in~\citet{zhang}, if we have independent
measurements of $M$, $d$, and $i$, the analysis of the disk's thermal spectrum 
can provide an estimate of $a_*$ and $\dot{M}$. This approach has been 
extensively discussed in the literature, its assumptions and limitations well
investigated and tested at both theoretical and observational level, and up to now 
the technique has been used to estimate the spin parameter of
about 10 stellar-mass BH candidates~\citep{li05,revcfm,revcfm2}. 
The continuum-fitting method cannot be used in the case of
supermassive objects because the disk's temperature goes like $M^{-0.25}$,
and for $M \sim 10^5 - 10^9$~$M_\odot$ the peak falls in the optical/UV range,
where dust absorption prevents good observations. The analysis of the disk's 
thermal spectrum can be naturally extended to non-Kerr spacetimes~\citep{cfm1,cfm2}.

In the present paper, we will use the code described in~\citet{cfm2} to constrain
the spin parameter and the deformation parameter of the 10~stellar-mass BH candidates
in X-ray binary systems for which the continuum-fitting method has already been
used by other authors to get an estimate of the spin parameter under the 
assumption that these objects are Kerr BHs. Our code is based on a ray-tracing
approach and includes all the special and general relativistic effects. The natural 
way to obtain these constraints would be to start from the raw data and repeat
the complete data analysis for each object, with the sole difference that the
thermal spectrum is now computed in a more general metric that includes the
Kerr solution as special case. While this would be surely the correct way to 
proceed, here we will employ a simplified analysis in which, significantly 
reducing the time of the analysis as well as the complications related to a large 
number of systematic effects, we will not lose any important information. The
idea is to compare the Kerr spectra with the spin parameters reported in the
literature with the spectra computed in our generic spacetime with a possible 
non-vanishing deformation parameter. The approach works because
the thermal spectrum of a thin disk around a non-Kerr BH with a certain spin
and deformation parameters is extremely similar to the one around a Kerr BH
with different spin and that remains true even for quite large deviations from
the Kerr solution. In other words, it is impossible to distinguish
a Kerr BH and a non-Kerr BH from the sole analysis of the disk's thermal 
spectrum~\citep{cfm2,bardeen} and therefore we can assume as reference spectrum 
the one of a Kerr BH with the spin reported in the literature. This approach significantly
simplify our job because we can just focus our attention on the role of the
spacetime metric, assuming that all the astrophysical effects and the 
instrumental issues have been already properly taken into account in the
previous studies. Our work is the first step of a project that aims at testing the
Kerr nature of BH candidates: the measurements reported in the present 
paper will be combined with other observations to try to break the degeneracy
between the spin and the deformation parameters. We note that
the use of the measurements from the continuum-fitting method and their
combinations with other observations have already been studied in previous
papers~\citep{jet1,jet2,qpo1}. However, those constraints 
were obtained with some simplifications that, as we will show 
here, provide the correct measurement in the limit of slow-rotating BHs 
and small inclination angles. 
Here we will remove such a set of simplifications.

The content of the paper is as follows. In Section~\ref{s-cfm}, we review the
calculations of the thermal spectrum of geometrically thin and optically thick
accretion disks around generic stationary and axisymmetric BHs. In 
Section~\ref{s-con}, we present our approach and, 
in Section~\ref{s-con-2}, we find the constraints on the spin 
parameter--deformation parameter
plane for each of the 10~stellar-mass BH
candidates for which a spin measurement from the disk's thermal
spectrum has been reported in the literature. 
Summary and conclusions are reported in Section~\ref{s-sc}. 
Throughout the paper, we use units in which $G_{\rm N} = c = 1$, unless 
stated otherwise.



\section{Thermal spectrum of thin disks}\label{s-cfm}

The calculation of the thermal spectrum of a thin disk around a BH consists
of two parts: the calculation of the disk structure and the one of the propagation
of the radiation from the disk to the observer, see e.g.~\citet{li05} and references 
therein. For the first part concerning the disk structure, we adopt the
Novikon-Thorne model~\citep{nt73,pt74}, which is the relativistic generalization 
of the Shakura-Sunyaev one~\citep{ss73} and describes geometrically thin
and optically thick accretion disks in generic stationary, axisymmetric, and 
asymptotically flat spacetimes. The model relies on a set of assumptions, 
whose validity has been tested by a number of theoretical and observational
studies~\citep{revcfm2}. These assumptions are thought to hold in the case
of geometrically thin and optically thick accretion disks, and the observational
criterion to select sources with similar disks is that the accretion luminosity
must be in the range $\sim 8-20$\% the Eddington luminosity. The disk is
supposed to be on the equatorial plane and the particles of the disk move 
on nearly geodesic circular orbits. From the conservation laws for the 
rest-mass, the energy, and the angular momentum, one obtains the equations
governing the radial structure of the disk~\citep{pt74}. The mass accretion rate
$\dot{M}$ is independent of the time and the radial coordinate, and the 
time-averaged energy flux is
\be
\mathcal{F}(r) &=& \frac{\dot{M}}{4\pi M^2} F(r) \, , 
\ee
where 
\be
F(r) = - \frac{\partial_r \Omega}{\left(E 
- \Omega L_z \right)^2} \frac{M^2}{\sqrt{-G}}
\int_{r_{\rm in}}^{r} \left(E - \Omega L_z \right) 
\left(\partial_\rho  L_z \right) \, d\rho \, .
\ee
$\Omega$, $E$, and $L_z$ are, respectively, the angular velocity 
$d\phi/dt$ of equatorial circular geodesics, the conserved specific energy, 
and the conserved axial component of the specific angular momentum.
$G = -\alpha^2 g_{rr} g_{\phi\phi}$ is the determinant of the near equatorial 
plane metric in cylindrical coordinates, and $\alpha^2 = g_{t\phi}^2/g_{\phi\phi} - g_{tt}$ 
is the lapse function. $r_{\rm in}$ is the radial coordinate of the inner edge
of the accretion disk and the key-point for what follows is that $r_{\rm in}$
is assumed to coincide with the innermost stable circular orbit (ISCO). 
The calculation of $\Omega$, $E$, and $L_z$ in a generic stationary and 
axisymmetric spacetime can be found, for instance, in Appendix~B of~\citet{cfm1}.

As the disk is in thermal equilibrium, it is possible to define an effective 
temperature $T_{\rm eff} (r) = [\mathcal{F}(r)/\sigma]^{1/4}$, where $\sigma$ is the 
Stefan-Boltzmann constant. In the case of stellar-mass BH candidates, the 
temperature of the disk can be at the level of $10^7$~K near its inner edge 
and therefore non-thermal effects have to be taken into account. That can be 
done by introducing the color temperature $T_{\rm col} (r) = f_{\rm col} T_{\rm eff}$, 
where $f_{\rm col}$ is usually referred as color factor or hardening factor, 
and the local specific intensity of the radiation emitted by the disk can be 
written as
\be\label{eq-i-bb}
I_{\rm e}(\nu_{\rm e}) = \frac{2 h \nu^3_{\rm e}}{c^2} \frac{1}{f_{\rm col}^4} 
\frac{\Upsilon}{\exp\left(\frac{h \nu_{\rm e}}{k_{\rm B} T_{\rm col}}\right) - 1} \, .
\ee
Here, $h$, $c$, and $k_{\rm B}$ are, respectively, the Planck constant, the speed of 
light, and the Boltzmann constant. $\nu_{\rm e}$ is the photon frequency in the 
rest frame of the emitter. $\Upsilon$ is a function of $\xi$, the angle between the 
wavevector of the photon emitted by the disk and the normal of the disk surface.
In the case of isotropic emission, $\Upsilon = 1$, while for limb-darkened emission 
we have $\Upsilon = \frac{1}{2} + \frac{3}{4} \cos\xi$.

The second part of the calculation of the thermal spectrum is related to the
propagation of the radiation from the disk to the observer. The image plane of
the observer has Cartesian coordinates $(X,Y)$. The photons emitted from 
the disk and detected by the observer have 3-momentum perpendicular to the 
image plane. The photon trajectory can be integrated backwards in time from
the image plane to the point of the emission on the accretion disk. In this 
way, one finds the radial coordinate $r_{\rm e}$ of the photon emission on the
equatorial plane and the angle $\xi$ between the wavevector of the photon 
and the normal of the disk surface. The image plane of the observer is divided 
into a number of small elements, and the ray-tracing procedure provides the 
observed flux density from each element. Integrating over all these elements, 
one finds the total observed flux density of the disk. The photon flux number 
density is
\be\label{eq-n2}
N_{E_{\rm obs}} =
\frac{1}{E_{\rm obs}} \int I_{\rm obs}(\nu) d \Omega_{\rm obs} = 
\frac{1}{E_{\rm obs}} \int g^3 I_{\rm e}(\nu_{\rm e}) d \Omega_{\rm obs} = 
A_1 \left(\frac{E_{\rm obs}}{\rm keV}\right)^2
\int \frac{1}{M^2} \frac{\Upsilon dXdY}{\exp\left[\frac{A_2}{g F^{1/4}} 
\left(\frac{E_{\rm obs}}{\rm keV}\right)\right] - 1} \, .
\ee
$E_{\rm obs}$, $I_{\rm obs}$, and $\nu$ are, respectively, the photon energy,
the specific intensity of the radiation, and the photon frequency measured by 
the distant observer. $I_{\rm e}(\nu_{\rm e})/\nu_{\rm e}^3 = I_{\rm obs} 
(\nu)/\nu^3$ is a consequence of the Liouville theorem. $d\Omega_{\rm obs} 
= dX dY / d^2$ is the element of the solid angle subtended by the image of the 
disk on the observer's sky, $d$ is the distance of the source, and $A_1$ and 
$A_2$ are given by
\be
A_1 &=&  
\frac{2 \left({\rm keV}\right)^2}{f_{\rm col}^4} 
\left(\frac{G_{\rm N} M}{c^3 h d}\right)^2 =  
\frac{0.07205}{f_{\rm col}^4} 
\left(\frac{M}{M_\odot}\right)^2 
\left(\frac{\rm kpc}{d}\right)^2 \, 
{\rm \gamma \, keV^{-1} \, cm^{-2} \, s^{-1}} \, , \nonumber\\
A_2 &=&  
\left(\frac{\rm keV}{k_{\rm B} f_{\rm col}}\right) 
\left(\frac{G_{\rm N} M}{c^3}\right)^{1/2}
\left(\frac{4 \pi \sigma}{\dot{M}}\right)^{1/4} = 
\frac{0.1331}{f_{\rm col}} 
\left(\frac{\rm 10^{18} \, g \, s^{-1}}{\dot{M}}\right)^{1/4}
\left(\frac{M}{M_\odot}\right)^{1/2} \, .
\ee
$g$ is the redshift factor
\be
g = \frac{\nu}{\nu_{\rm e}} = 
\frac{k_\mu u^\mu_{\rm o}}{k_\mu u^\mu_{\rm e}} \,,
\ee
where $u^\mu_{\rm o} = (1,0,0,0)$ is the 4-velocity of the observer and 
$u^\mu_{\rm e}=(u^t_{\rm e},0,0,\Omega u^t_{\rm e})$ is the 4-velocity of the emitter. 
From the normalization condition $g_{\mu\nu}u^\mu_{\rm e} u^\nu_{\rm e} = -1$, 
one finds $u^t_{\rm e}$ as a function of $r_{\rm e}$
\be
u^t_{\rm e} = \frac{1}{\sqrt{-g_{tt} - 2g_{t\phi}\Omega 
- g_{\phi\phi}\Omega^2}} \, .
\ee
Since the spacetime is stationary and axisymmetric, $k_\phi/k_t = \lambda$ is
a constant along the photon path and it can be calculated from the photon 
initial conditions on the image plane of the observer, where the spacetime can 
be assumed flat. One eventually finds the redshift factor as a function of $X$
and $Y$
\be
g = \frac{\sqrt{-g_{tt} - 2g_{t\phi}
\Omega - g_{\phi\phi}\Omega^2}}{1 + \lambda \Omega} \, .
\ee
The relativistic effects of Doppler boost and gravitational redshift are entirely
encoded in $g$. The ray-tracing calculation adds the effect of light bending.



\section{Testing the nature of black hole candidates} \label{s-con}

The formalism reviewed in the previous section is very general, in the sense that it
can be applied for any stationary, axisymmetric, and asymptotically flat spacetime.
In the special case of the Kerr background, the ray-tracing part of the calculation
of the photon trajectories from the disk to the observer's image plane can be 
significantly simplified, because the spacetime is of Petrov type~D and in 
Boyer-Lindquist coordinates the photon equations of motion are separable and 
of first order. This is not true in the general case, and our code solves the
second order photon geodesic equations. As background geometry, here we
adopt the Johannsen-Psaltis metric. In Boyer-Lindquist coordinates, the line 
element is~\citep{metric}
\be\label{eq-jp}
ds^2 &=& - \left(1 - \frac{2 M r}{\Sigma}\right) (1 + h) \, dt^2
- \frac{4 a M r \sin^2\theta}{\Sigma} (1 + h) \, dt \, d\phi
+ \frac{\Sigma (1 + h)}{\Delta + a^2 h \sin^2\theta } \, dr^2 + \nonumber\\
&& + \Sigma \, d\theta^2
+ \left[\sin^2\theta \left(r^2 + a^2 + \frac{2 a^2 M r \sin^2\theta}{\Sigma} \right) 
+ \frac{a^2 (\Sigma + 2 M r) \sin^4\theta}{\Sigma} h \right] d\phi^2 \, , 
\ee
where
\be
\Sigma &=& r^2 + a^2 \cos^2\theta \, , \nonumber\\
\Delta &=& r^2 - 2 M r + a^2 \, , \nonumber\\
h &=& \sum_{k = 0}^{\infty} \left(\epsilon_{2k} 
+ \frac{M r}{\Sigma} \epsilon_{2k+1} \right)
\left(\frac{M^2}{\Sigma}\right)^k \, .
\ee
The Johannsen-Psaltis metric has an infinite number of deformation parameters 
$\epsilon_i$ and it includes the Kerr solution as the special case in which all the 
deformation parameters vanish. 
The correct Newtonian limit is recovered for $\epsilon_0 = 0$. Solar System 
experiments require $\epsilon_1$, $\epsilon_2 \ll 1$ (assuming the validity of 
the Birkhoff theorem)~\citep{jp-vc}. In the rest of this paper, we restrict our attention to 
the deformation parameter $\epsilon_3$ and we set to zero all the others. 
All the deformation parameters have a similar impact on the 
geometry of the spacetime~\citep{agn}. 
For $\epsilon_i > 0$, the BH is more oblate than its Kerr counterpart with the 
same spin and the gravitational force on the equatorial plane is weaker. For
$\epsilon_i < 0$, the BH is more prolate than the Kerr one and the gravitational 
force on the equatorial plane is stronger. 
Here, we consider a single deformation parameter because this is 
the simplest case. While this choice cannot address generic deviations from the
Kerr solution, with current observations it is already difficult to get an estimate 
of one deformation parameter.

With the metric in Eq.~(\ref{eq-jp}), we can compute the expected thermal spectrum
of thin accretion disks and compare the predictions with observational data. 
The result is a constraint on the spin parameter--deformation parameter
plane. Let us note that, strictly speaking, this approach can only test the Kerr geometry 
around a BH candidate and does not test the validity of the Einstein equations. Indeed, while 
the Kerr metric is the only vacuum BH solution in 4-dimensional general relativity, it is a
solution even in other theories of gravity~\citep{k1}. 
The thermal spectrum of a thin disk is determined by the geodesic 
motion of the gas in the disk and of the photons from the disk to the observer.
Our approach cannot distinguish a Kerr BH of general relativity from a Kerr BH in 
another theory of gravity because the geodesic equations are the same in the two 
scenarios.

Tab.~\ref{tab} shows the spin measurements from the continuum-fitting method
reported in the literature and obtained assuming the Kerr nature of BH candidates. 
The first column reports the name of the binary system, the second column shows the 
mass measurement of the BH candidates, the third column is for the estimate of the 
inclination angle of the disk, the forth column reports the spin measurement, and 
the fifth column is for the reference of the spin measurement. The BH mass $M$ 
and the inclination angle of the disk $i$ are usually inferred with dynamical models 
and optical observations, as they are input parameters in the continuum-fitting 
approach. In some cases, the angle $i$ is estimated from the orientation of the BH 
jet, assuming that it is parallel to the BH spin and perpendicular to the accretion 
disk. See the original papers for more details and~\citet{revcfm,revcfm2} for a 
general discussion. In what follows, we only consider the first 10~BH candidates
in this table, while we neglect the last 3. The constraint on the spin parameter of
the BH candidate in GX~339-4 is 
very weak and it could only rule out half part of the spin 
parameter--deformation parameter plane.
The two recent studies of GS~1124-683 and of the microquasar
in M31 seem to indicate that the accretion disk of these objects is counterrotating, but
in this case the applicability of the continuum-fitting method is more questionable.
So we prefer to focus our attention only on those measurements that are supposed 
to be the most robust and reliable.

\begin{table} 
\centering
\begin{tabular}{ccccccccc}
\hline \hline
BH Binary & \hspace{0.1cm} & $M/M_\odot$ & \hspace{0.1cm} & $i$ & \hspace{0.1cm} & $a_*$ & \hspace{0.1cm} & Reference \\
\hline \hline
GRO~J1655-40 && $6.30 \pm 0.27$ && $70.2^\circ \pm 1.2^\circ$ && $0.70 \pm 0.10$ && \citet{sh06} \\
4U~1543-47 && $9.4 \pm 1.0$ && $20.7^\circ \pm 1.5^\circ$ && $0.80 \pm 0.10$ && \citet{sh06} \\
GRS~1915+105 && $14.0 \pm 4.4$ && $66^\circ \pm 2^\circ$ && $> 0.98$ && \citet{1915} \\
M33~X-7 && $15.65 \pm 1.45$ && $74.6^\circ \pm 1.0^\circ$ && $0.84 \pm 0.05$ && \citet{liu08,liu10} \\
LMC~X-1 && $10.91 \pm 1.54$ && $36.38^\circ \pm 2.02^\circ$ && $0.92^{+0.05}_{-0.07}$ && \citet{lmcx1} \\
A0620-00 && $6.61 \pm 0.25$ && $51.0^\circ \pm 0.9^\circ$ && $0.12 \pm 0.19$ && \citet{62} \\
XTE~J1550-564 && $9.10 \pm 0.61$ && $74.7^\circ \pm 3.8^\circ$ && $0.34^{+0.20}_{-0.28}$ && \citet{xte} \\
Cygnus~X-1 && $14.8 \pm 1.0$ && $27.1^\circ \pm 0.8^\circ$ && $> 0.98$ && \citet{cyg1,cyg2} \\
H1743-322 && $\sim 10$ && $75^\circ \pm 3^\circ$ && $0.2 \pm 0.3$ && \citet{h1743} \\
LMC~X-3 && $6.95 \pm 0.33$ && $69.6^\circ \pm 0.6^\circ$ && $0.21 \pm 0.12$ && \citet{lmcx3} \\
\hline
GX~339-4 && $5.8 - 15$ && $20^\circ - 70^\circ$ && $< 0.9$ && \citet{gx339} \\
GS~1124-683 && $7.24 \pm 0.70$ && $54^\circ \pm 1.5^\circ$ && $< -0.2$ && \citet{gs1124} \\
M31~microquasar && $\sim 10$ && $\sim 32^\circ$, $\sim 39^\circ$ && $< -0.2$ && \citet{m31} \\
\hline \hline
\end{tabular}
\caption{Continuum-fitting measurements of the spin parameter of stellar-mass BH 
candidates reported in the literature under the assumption of the Kerr background. 
See the references in the last column for more details. \label{tab}}
\end{table}

In our non-Kerr model, the thermal spectrum of thin disks depends 
on 3 free parameters, namely $a_*$, $\epsilon_3$, and $\dot{M}$, because with the
continuum-fitting method $M$, $d$, and $i$ are to be determined by independent 
measurements. One can thus compute a number of spectra and fit the data to infer
$a_*$, $\epsilon_3$, and $\dot{M}$. In this paper, we use instead an approach that 
exploits the fundamental degeneracy between $a_*$ and $\epsilon_3$. While the 
procedure significantly simplify the analysis, we argue (see below) that the final result 
is very similar, especially in the cases of slow-rotating objects, which show a true 
degeneracy between $a_*$ and $\epsilon_3$. We replace the observational data
with the theoretical spectrum of a Kerr BH with spin parameter given by the measurement
shown in Tab.~\ref{tab}. We compute the $\chi^2$ by comparing this 
reference spectrum with the one of a Johannsen-Psaltis BH with spin parameter 
$a_*$, deformation parameter $\epsilon_3$, and mass accretion rate $\dot{M}$:
\be
\chi^2 (a_*, \epsilon_3, \dot{M}) =
\sum_{i=1}^n \frac{\left[ N_i (a_*, \epsilon_3, \dot{M}) 
- N_i' (a_*', \epsilon_3', \dot{M}') \right]^2}{\sigma^2_i} \, ,
\ee
where the summation is performed over $n$ sampling energies $E_i$, $N_i$ and
$N_i'$ are, respectively, the photon fluxes in the energy bin $[E_i, E_i + \Delta E]$ for
the non-Kerr model and the reference Kerr model. $a_*'$, $\epsilon_3' = 0$, and 
$\dot{M}'$ are the spin parameter, the vanishing deformation parameter, and the 
mass accretion rate of the reference Kerr model. The error $\sigma_i$ is assumed to
be proportional to the mean value of $N_i$ and $N_i'$, but our results will be 
independent of the constant of proportionality. This is a simple method to quantify
the level of correspondence between the spectra of the two models~\citep{v2-jjc}.  
We note that we should have assumed $\sigma_i$ proportional to the
square root of the mean value of $N_i$ and $N_i'$ if either a measurement was 
obtained using some sort of photon counting instrument or we are making predictions 
about what a hypothetical detector's error would be. Here we are comparing the 
correspondence of two models in model-space; that is, something which has nothing 
to do with a physical detector and it makes sense to assume that the error is proportional 
to the mean value of $N_i$ and $N_i'$.

The mass $M$ and the distance $d$ only alter the normalization of the photon flux; 
here we use the same values in the two models (we basically assume that the 
dynamical measurements are correct) and they do not have any effect on our final 
constraints. The value of the inclination angle $i$ is instead important, because it 
has an impact on the effect of light bending, and we use the measurements shown 
in the third column of Tab.~\ref{tab} for any object. Concerning the mass accretion 
rate, within the continuum-fitting method it is an output parameter to be inferred
during the fitting procedure. We cannot use the mass accretion rate found in the
measurements reported in the literature, because its estimate is correlated to the 
one of the spin parameter [see Fig.~4a in~\citet{lmcx1}]. The measurement of the 
accretion luminosity $L_{\rm acc} = \eta \dot{M}$, where $\eta$ is the radiative 
efficiency, is instead not correlated to the measurement of $a_*$ [see Fig.~4b 
in~\citet{lmcx1}]. In our analysis we thus use a constant accretion luminosity and
$\dot{M}$ is given by:
\be\label{eq-mdot}
\dot{M} = \frac{\eta' \dot{M}'}{\eta} \, ,
\ee 
where $\eta = \eta(a_*,\epsilon_3)$ and $\eta'$ are, respectively, the radiative 
efficiency in the non-Kerr and reference Kerr models, namely $\eta = 1 - E_{\rm ISCO}
(a_*,\epsilon_3)$ and $\eta' = 1 - E_{\rm ISCO}(a_*')$, where $E_{\rm ISCO}$ is 
the specific energy at the ISCO radius and depends on the background metric.
The exact value of the accretion luminosity does not affect the final results, so the
important point is that we impose that $\dot{M}$ scales as the inverse of $\eta$.

\begin{figure}
\begin{center}
\includegraphics[height=8cm]{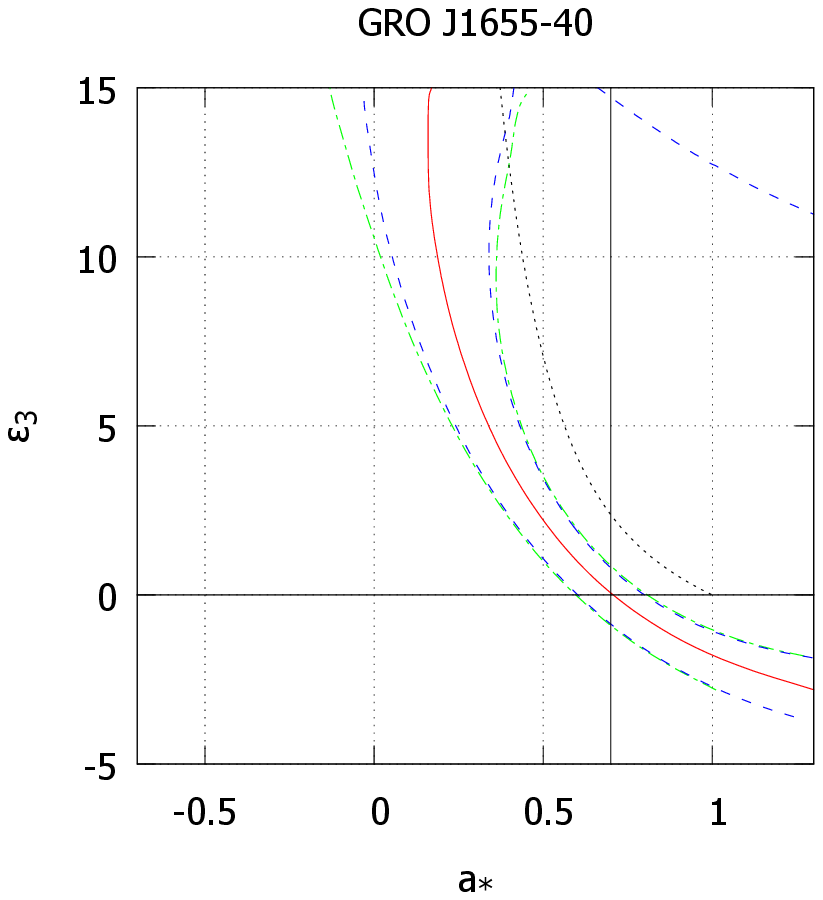}
\includegraphics[height=8cm]{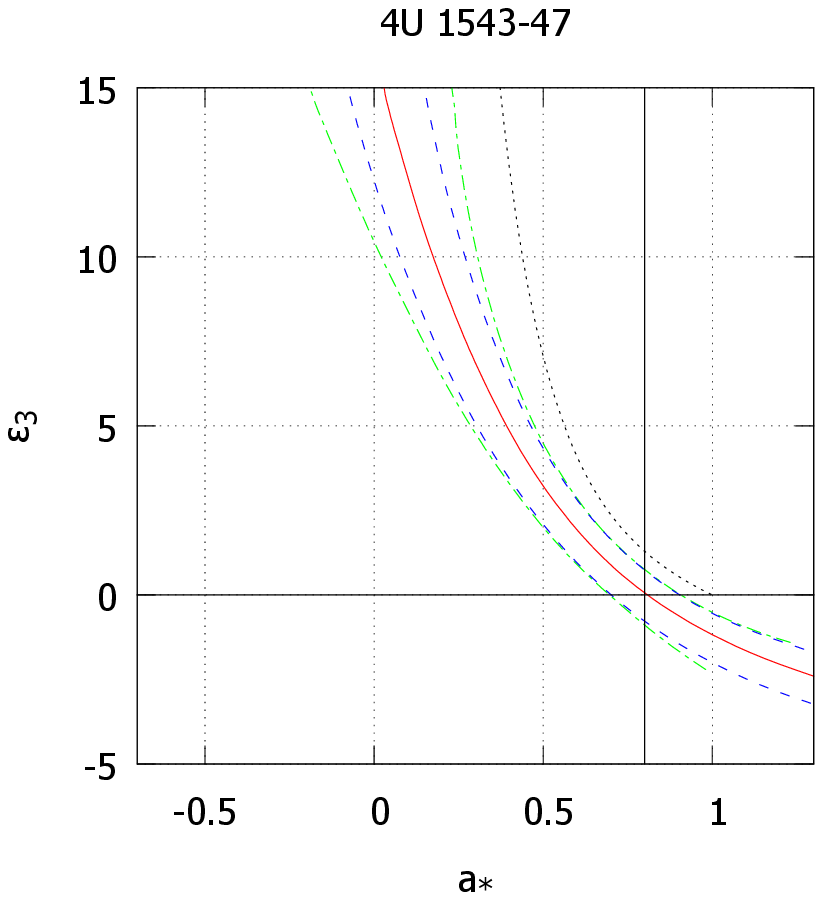}
\end{center}
\caption{Disk's thermal spectrum constraints on possible deviations from the 
Kerr geometry in the spacetime around the BH candidates in GRO~J1655-40 
(left panel) and 4U~1543-47 (right panel). Here and in the other plots of this 
paper, the red-solid curve is $\chi^2_{\rm min}(\epsilon_3)$, the blue-dashed
curves are $\sigma_-$ and $\sigma_+$, respectively on the left and right side of
the red-solid line, while the green-dashed-dotted curves are $\chi^2_{\rm min}(\epsilon_3) 
+ \sigma_-$ and $\chi^2_{\rm min}(\epsilon_3) + \sigma_+$, respectively on 
the left and right side of the red-solid line. See the text for more details. \label{fig1}}
\end{figure}

\begin{figure}
\begin{center}
\includegraphics[height=8cm]{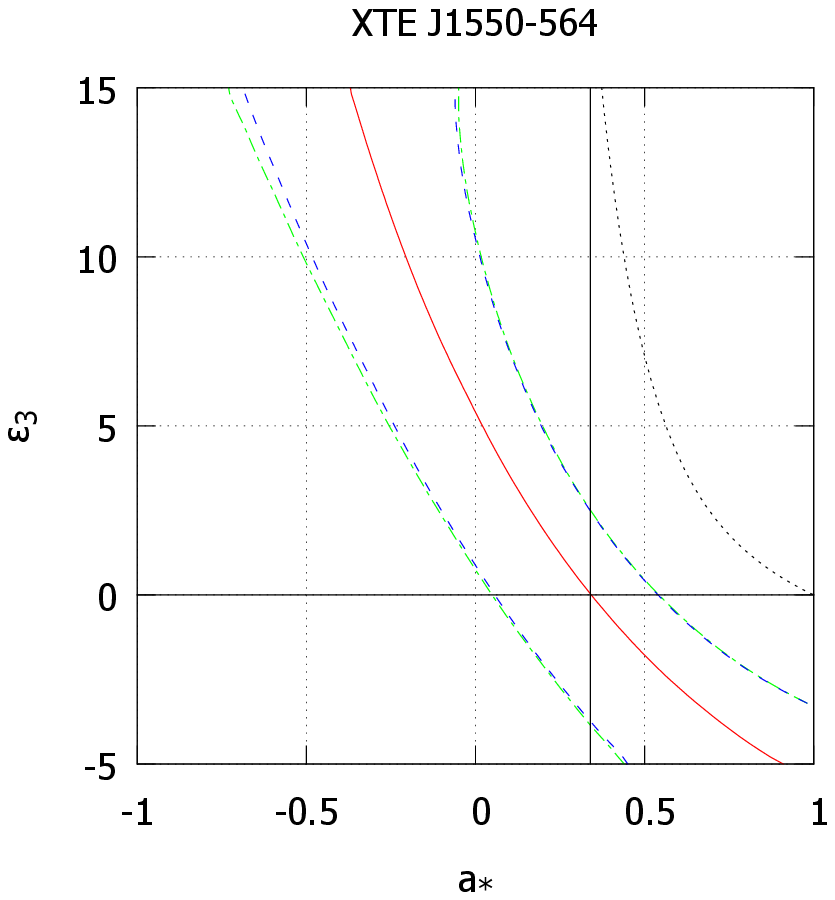}
\includegraphics[height=8cm]{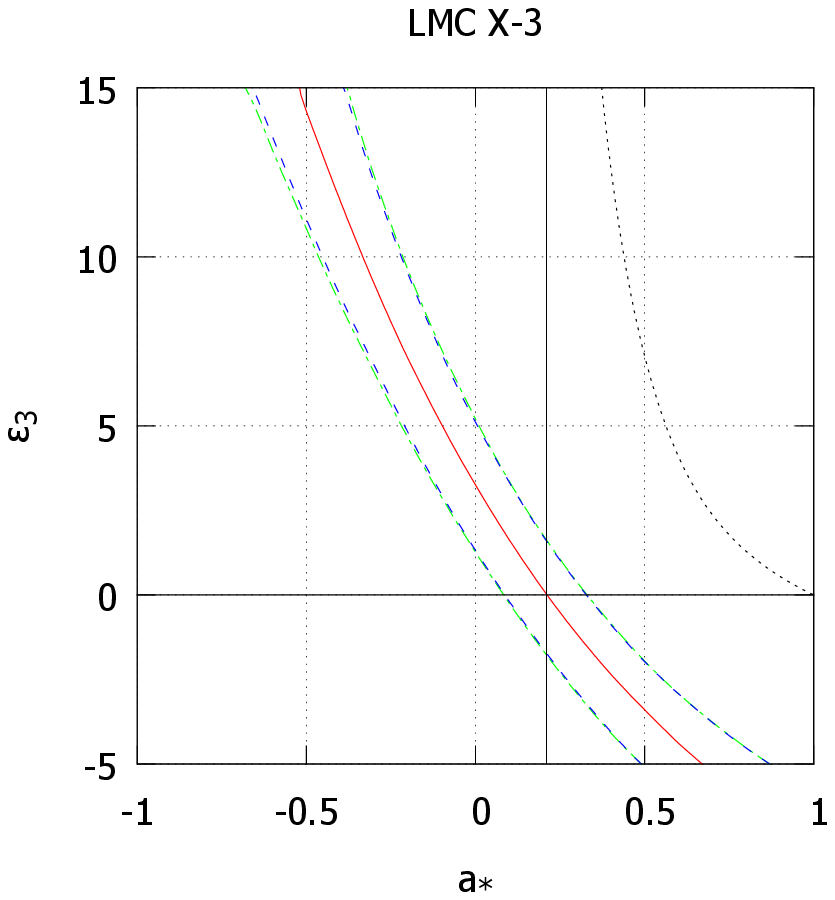}\\
\includegraphics[height=8cm]{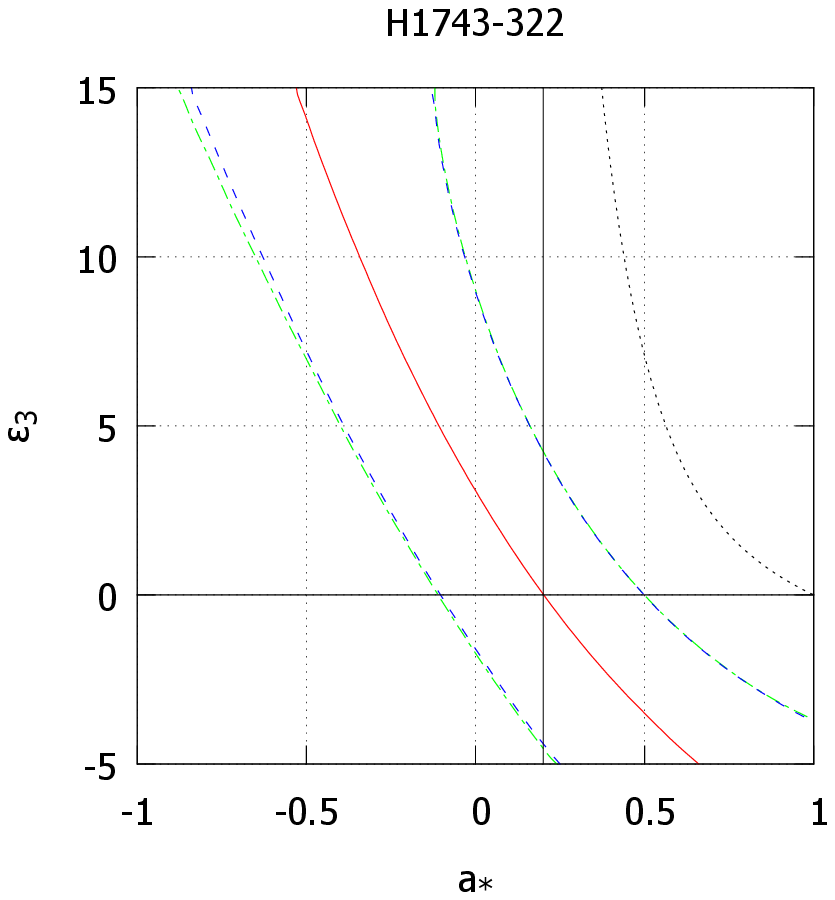}
\includegraphics[height=8cm]{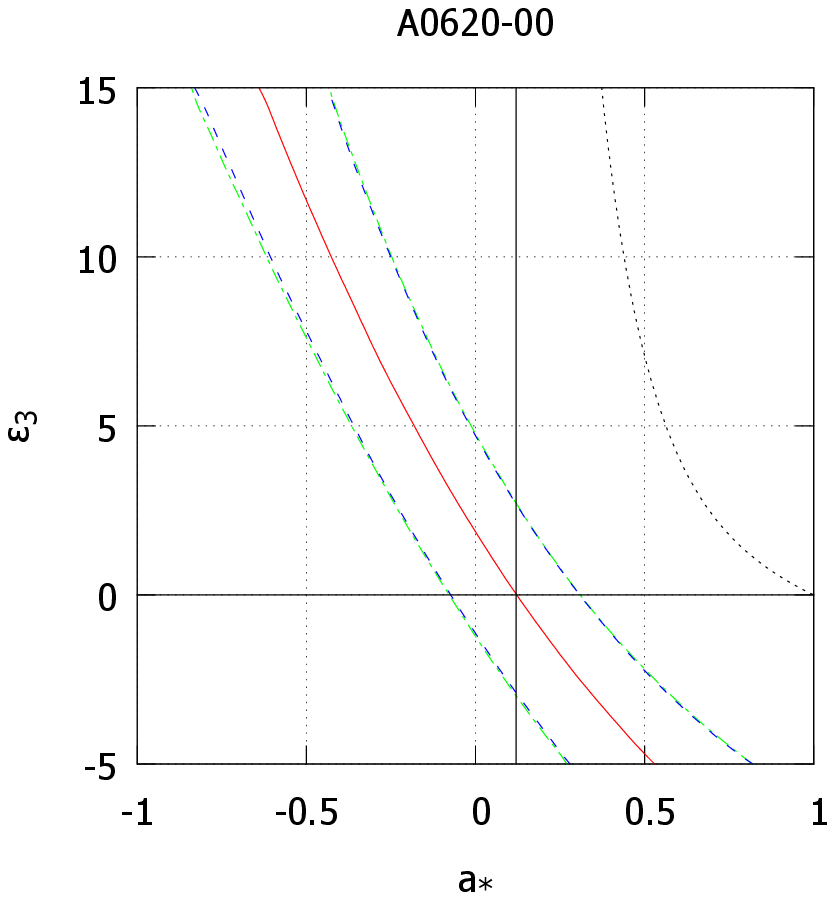}
\end{center}
\caption{As in Fig.~\ref{fig1} for the BH candidates in XTE~J1550-564 (top left 
panel), LMC~X-3 (top right panel), H1743-322 (bottom left panel), and A0620-00 
(bottom right panel). See the text for more details. \label{fig2}}
\end{figure}

For any value of $\epsilon_3$, $\chi^2$ has one or more local minima, 
which can be indicated by $\chi^2_{\rm min}(\epsilon_3)$. For $\epsilon_3 = 0$, the 
minimum of $\chi^2$ is exactly zero, by definition, but it is usually very close to zero 
even for a non-vanishing $\epsilon_3$. Let us call $\sigma_-$ and $\sigma_+$ the 
values of $\chi^2$ for $\epsilon_3 = 0$ corresponding, respectively, to the 1-sigma 
lower and upper measurements of the BH spin parameter in Tab.~\ref{tab}. We can 
then define two kinds of constraints on the spin parameter--deformation parameter 
plane. If BH candidates are really Kerr BHs, the analysis of real data would be able to
rule out all the models with $\chi^2 > \sigma_-$ on the left of $\chi^2_{\rm min}(\epsilon_3)$ 
and all the models with $\chi^2 > \sigma_+$ on the right of $\chi^2_{\rm min}(\epsilon_3)$.
As already noted, in this way it is not important the constant of proportionality between
$\sigma_i$ and the mean value of $N_i$ and $N_i'$.
The second type of constraint is obtained by considering the region with 
$\chi^2 < \chi^2_{\rm min}(\epsilon_3) + \sigma_-$ on the left of 
$\chi^2_{\rm min}(\epsilon_3)$ and the region with
$\chi^2 < \chi^2_{\rm min}(\epsilon_3) + \sigma_+$ on the right of $\chi^2_{\rm min}(\epsilon_3)$.
In presence of a perfect degeneracy between the $a_*$ and $\epsilon_3$, the two
kinds of constraints would be equivalent, as $\chi^2_{\rm min}(\epsilon_3)$ should 
vanish. This is indeed what happens for slow-rotating objects within a good 
approximation. However, within our framework of the Johannsen-Psaltis metric with
deformation parameter $\epsilon_3$, such a degeneracy may not be realized
for fast-rotating objects and 
high values of $\epsilon_3$. In this case, it would make sense to 
consider the second type of constraint region, but actually one should be able 
to see it within a Kerr model, finding that the minimum of $\chi^2$ is much larger 
than the expected one. Within a Kerr model, this would be interpreted as a sign 
of over-fitting, over-confidence, or some other analysis error. In what follows, we 
show both the constraints: the second one can at least be used as a guide to figure 
out when $\chi^2_{\rm min}(\epsilon_3)$ is close to zero and when it is not.

\section{Results} \label{s-con-2}

\begin{figure}
\begin{center}
\includegraphics[height=8cm]{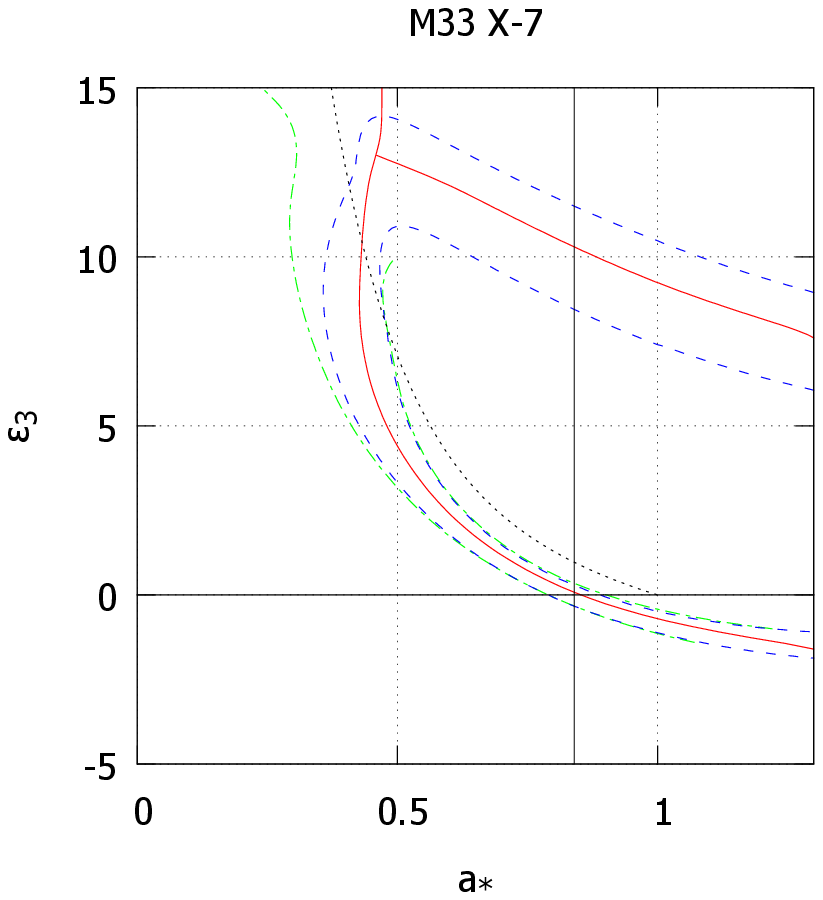}
\put(-180,90){Excluded Region}
\put(-100,115){Excluded Region}
\put(-90,185){Excluded Region}
\includegraphics[height=8cm]{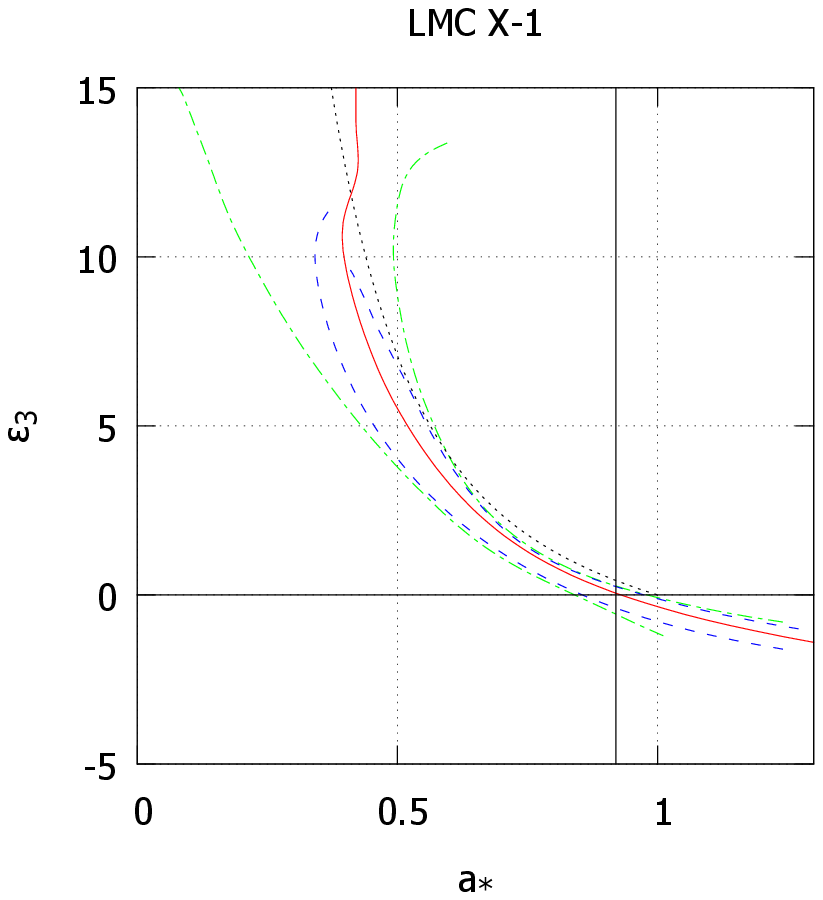}
\put(-180,90){Excluded Region}
\put(-100,140){Excluded Region}
\end{center}
\caption{As in Fig.~\ref{fig1} for the BH candidates in M33~X-7 (left 
panel) and LMC~X-1 (right panel). See the text for more details. \label{fig2b}}
\begin{center}
\includegraphics[height=8cm]{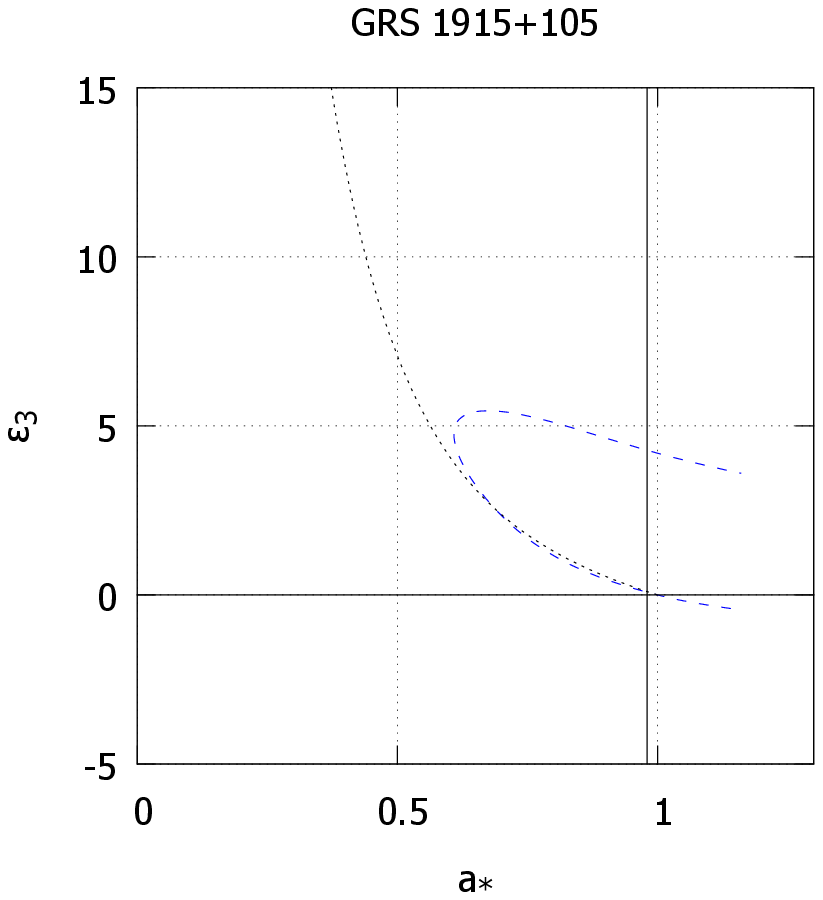}
\put(-90,97){Allowed Region}
\put(-160,140){Excluded Region}
\includegraphics[height=8cm]{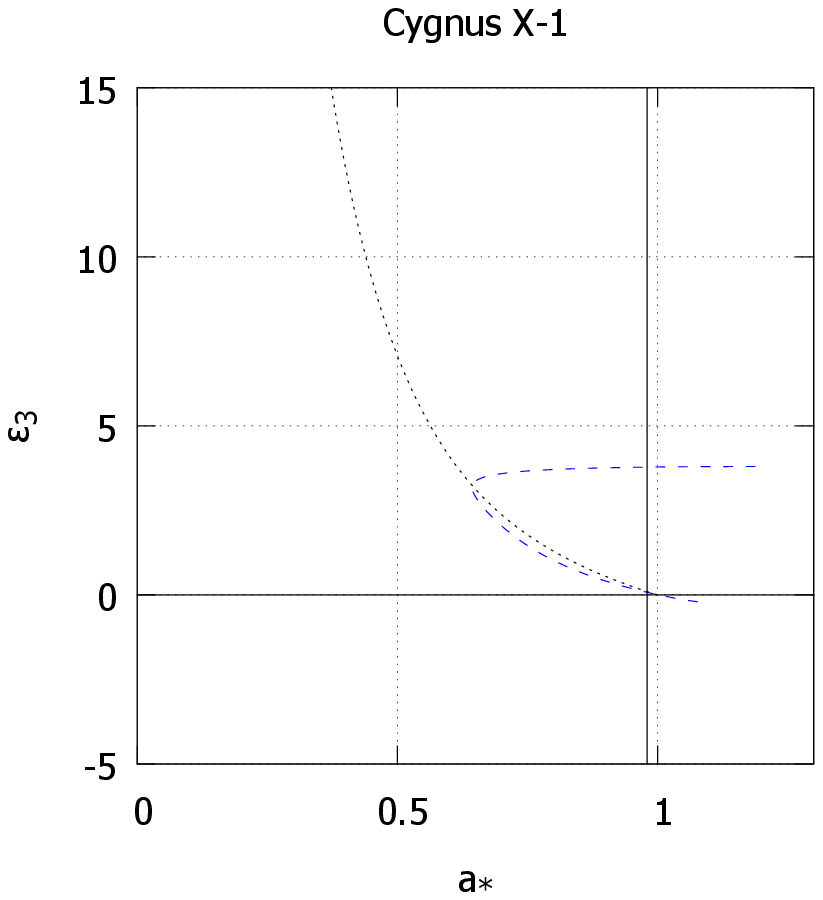}
\put(-90,97){Allowed Region}
\put(-160,140){Excluded Region}
\end{center}
\caption{As in Fig.~\ref{fig1} for the BH candidates in GRS~1915+105 (left 
panel) and Cygnus~X-1 (right panel). See the text for more details. \label{fig3}}
\end{figure}

Figs.~\ref{fig1}-\ref{fig3} show the constraints for the 10~stellar-mass BH 
candidates obtained with the procedure 
discussed in Section~\ref{s-con}. We note that a large region
of the spin parameter--deformation parameter plane describes spacetimes with naked
singularities [see Fig.~2 in~\citet{metric}]. 
In all our figures, the black dotted line starting from $a_* = 1$ and 
$\epsilon_3 = 0$ and moving to lower $a_*$ for $\epsilon_3 > 0$ is the boundary
separating spacetime with and without naked singularities. Spacetimes with naked 
singularities are on the right side of the black dotted line and in the Kerr limit 
$\epsilon_3 = 0$ we recover the well known fact that a naked singularity is present
for $a_* > 1$. For $\epsilon_3 < 0$, the BH horizon looks like a doughnut with a 
singularity at the center for $a_* > 1$~\citep{h-topo}. 
Here we have not automatically excluded such 
a region, because the continuum-fitting method can only test the spacetime geometry 
at radii larger than the ISCO and we have no information on the spacetime at smaller radii.
For instance, the metric in Eq.~(\ref{eq-jp}) may be the exterior solution of an
exotic object and be valid up to the radius of its surface, while at smaller radii we
could have an interior solution without singularities. We also note that in some plots the 
contours of $\chi^2$ end at some point. This is either because we are at the boundary of 
our plane and there are not enough calculated spectra, or because there are peculiar
features that make the constraint ambiguous (for instance, the green-dashed-dotted line in presence of two different local minima for a fixed $\epsilon_3$). In any case, this only
happens for a few cases and large deviations from Kerr, so it is not a problem for the
future applications of these constraints that we have in mind.

Concerning the reliability of the bounds found with this approach, 
we note that the thermal spectrum of thin disks has a very simple shape~\citep{rev2}. 
In most cases, there is a true degeneracy between $a_*$ and $\epsilon_3$. 
We thus believe that there is not the danger of erroneously weighting different 
portion of the spectrum, just because the same result should be generally found
from the analysis of every part. From this point of view, the only BH candidate that
might benefit of a complete data reanalysis is the object in LMC~X-1, for which
at high values of $\epsilon_3$ there is a departure from the spectrum obtained
in the Kerr background. The approach used here would provide instead more 
suspicious results in the case of the study of features with a more complicated 
structure, e.g. the iron K$\alpha$ line, because in that case different relativistic 
effects show up in different parts of the spectrum and thus it is important to
compare different parts of the spectrum with their actual uncertainty.

\subsection{GRS~J1655-40 and 4U~1543-47}

GRS~J1655-40 and 4U~1543-47 were the first two sources for which 
the continuum-fitting method was applied to estimate the spin parameters of
BH candidates~\citep{sh06} [previous attempts in~\citet{zhang} were only to 
illustrate the procedure, but serious measurements were not possible because of the
poor quality of the data and of systematic effects not fully under control]. Our
constraints on the spin parameter--deformation parameter plane 
are reported in Fig.~\ref{fig1}. The red-solid lines 
track the local minima of $\chi^2$, $\chi^2_{\rm min}(\epsilon_3)$. The blue-dashed
lines are for $\sigma_-$ and $\sigma_+$, respectively on the left and right side of
the red-solid lines. As already discussed, $\sigma_-$ and $\sigma_+$
correspond, respectively, to the 1-sigma lower and upper measurements
of the BH spin parameter in Tab.~\ref{tab}. For instance, in the case of GRO~J1655-40, 
$\sigma_- = \chi^2(a_*=0.6,\epsilon_3=0)$ and $\sigma_+ = \chi^2(a_*=0.8,\epsilon_3=0)$. 
The regions on the left side of the red-solid lines and between the red-solid lines 
and the blue-dashed lines include BHs with $0 < \chi^2 < \sigma_-$. The regions on
the right hand side of the red-solid lines and between the red-solid lines and the 
blue-dashed lines are for BHs with $0 < \chi^2 < \sigma_+$. Objects with $a_*$ and 
$\epsilon_3$ between the two blue-dashed lines are indistinguishable from a Kerr 
BH with the reference spin within the current uncertainties. The green-dashed-dotted 
lines are instead for $\chi^2_{\rm min}(\epsilon_3) + \sigma_-$ and 
$\chi^2_{\rm min}(\epsilon_3) + \sigma_+$, respectively on the left and right sides 
of the red-solid lines. Both for GRS~J1655-40 and 4U~1543-47 the difference
between the two kinds of constraints is small.

\begin{figure}
\begin{center}
\includegraphics[height=8cm]{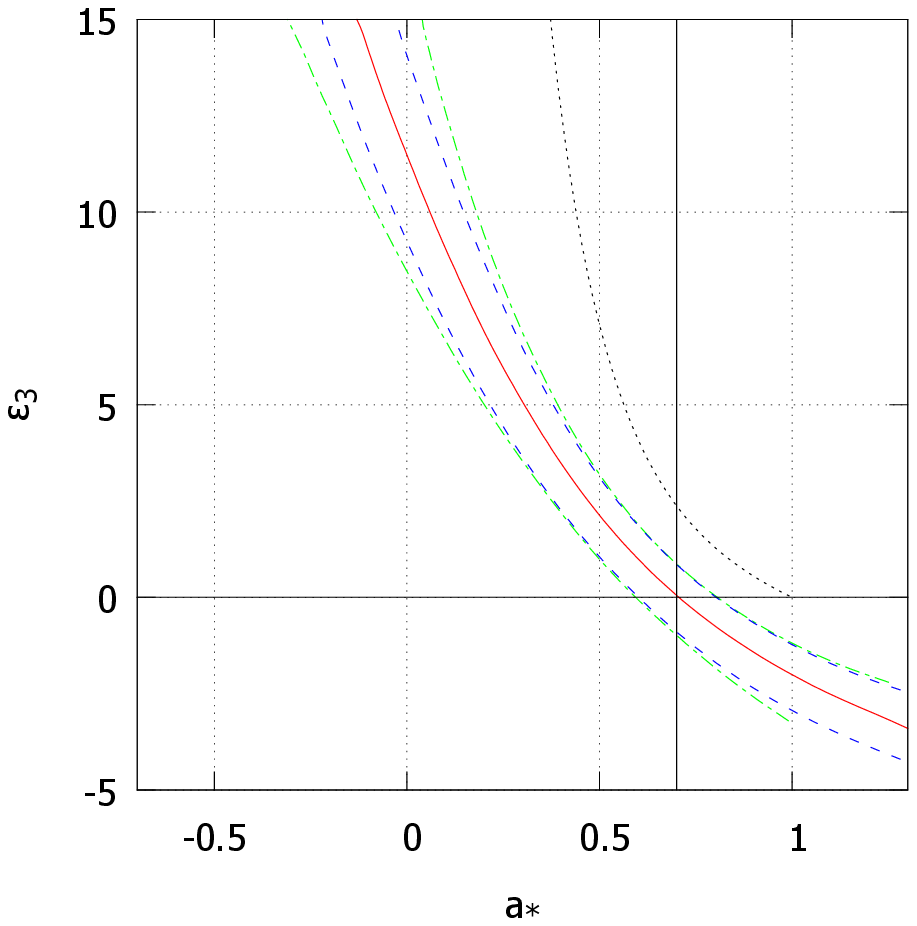} \\
\includegraphics[height=8cm]{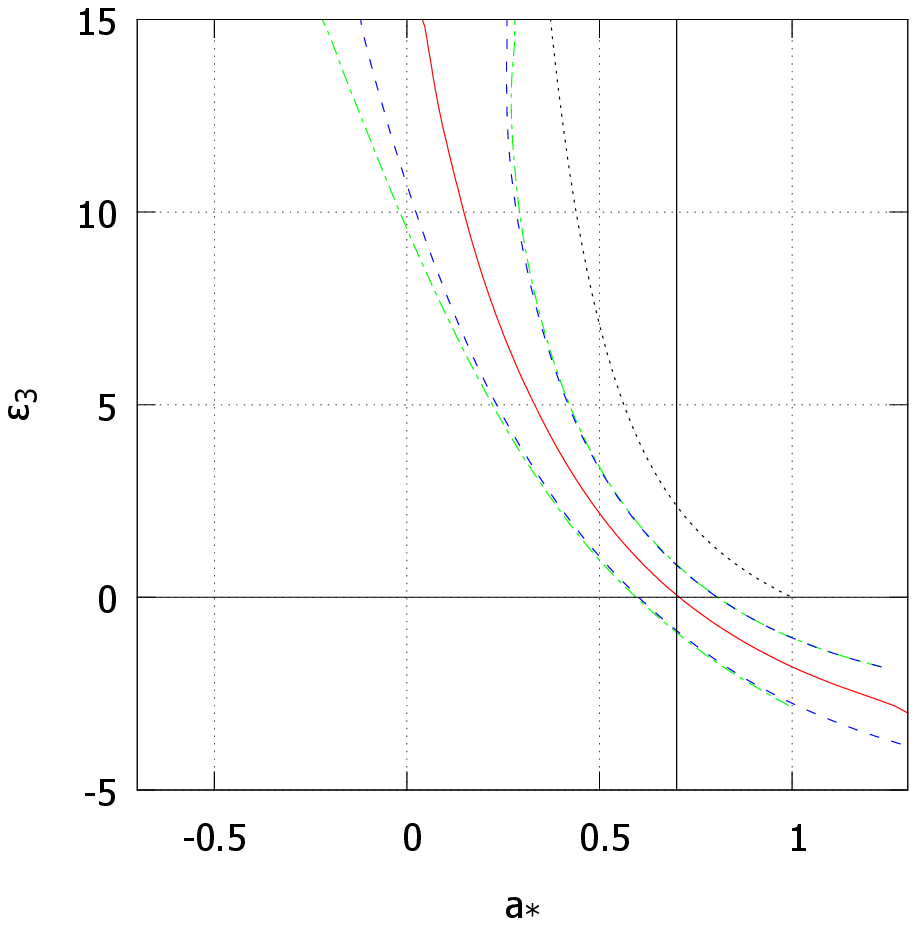}
\includegraphics[height=8cm]{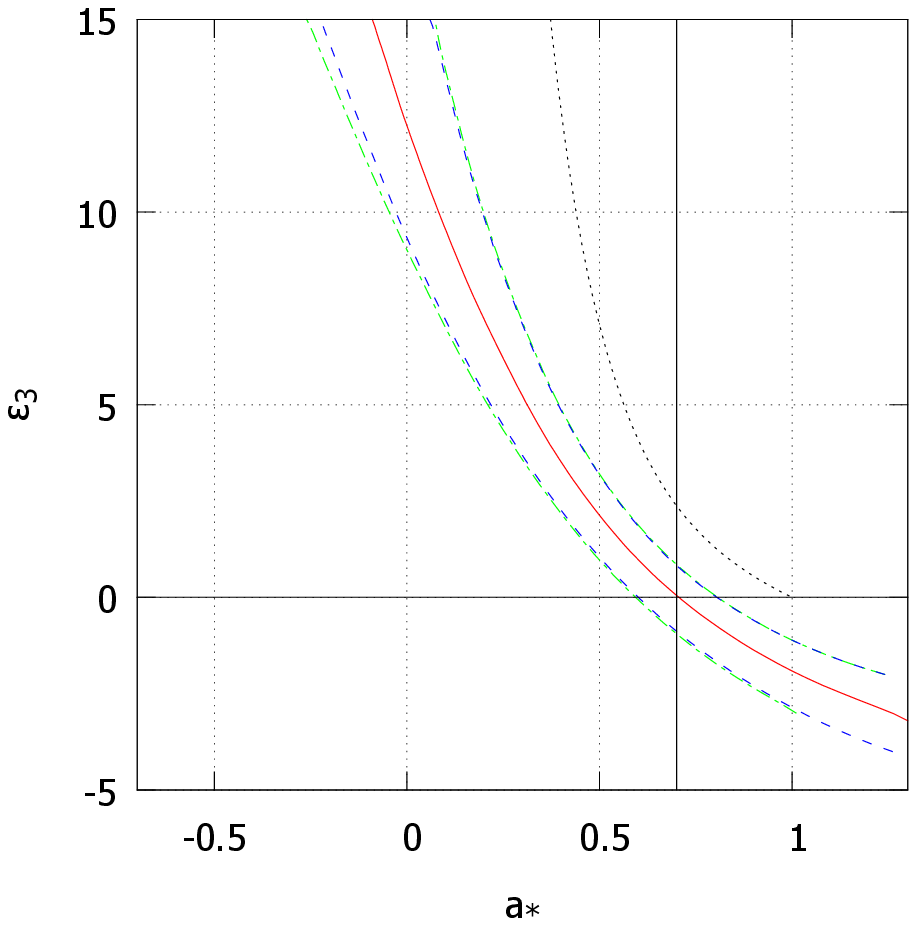}
\end{center}
\caption{Top panel: Disk's thermal spectrum constraints on possible deviations 
from the Kerr geometry in the spacetime around the BH candidate in 
GRO~J1655-40 that would have been obtained with an inclination angle 
$i = 20^\circ$. Bottom panels: as in the top panel, using a constant mass
accretion rate $\dot{M}=\dot{M}'$ instead of $\dot{M}=\eta' \dot{M}'/\eta$
in Eq.~(\ref{eq-mdot}) and an inclination angle $i = 70.2^\circ$ (left panel)
and $20^\circ$ (right panel). See the text for more details. \label{fig1b}}
\end{figure}

While the constraints on the spin parameter--deformation parameter 
plane primarily depend on the measurement of the spin in the Kerr background,
a minor role is played by the value of the disk's inclination angle with respect to our 
line of sight. 
The effect of $i$ becomes more pronounced for $\epsilon_3 > 10$: 
for high values of $\epsilon_3$, the preferred spin parameter $a_*$ is larger for 
higher $i$ than the one that would been obtained for lower $i$. To make more clear 
this statement, in the top panel of Fig.~\ref{fig1b} we show the continuum-fitting 
method constraints that would have been obtained for GRO~J1655-40 assuming an 
inclination angle $i = 20^\circ$ instead of $i = 70.2^\circ$. At $\epsilon_3 = 15$, 
the spin measurement for $i = 20^\circ$ is $a_* = -0.13^{+0.10}_{-0.18}$, while for 
$i = 70.2^\circ$ one finds $a_* = 0.17\pm 0.30$. For instance, the inclination angle 
is the key-point for the difference between the shape of the constraints for GRO~J1655-40 
and 4U~1543-47 in Fig.~\ref{fig1}. In the Kerr background, one finds respectively
$a_* = 0.7 \pm 0.1$ and $a_* = 0.8 \pm 0.1$; that is, 4U~1543-47 rotates faster 
than GRO~J1655-40. For high values of the deformation parameter, one finds 
instead that GRO~J1655-40 would rotate faster than 4U~1543-47.

Concerning the choice of the mass accretion rate, here we are
assuming the same accretion luminosity $L_{\rm acc} = \eta \dot{M}$ in all the 
models. As shown in~\citet{lmcx1}, the estimate of $a_*$ and $\dot{M}$ in the Kerr 
background are correlated, and therefore in our analysis we cannot assume a constant 
mass accretion rate. On the contrary, the estimate of the accretion luminosity is not 
correlated with the measurement of $a_*$. 
In Fig.~\ref{fig1b}, we show the constraints 
that would have been obtained for GRO~J1655-40 in the case of $\dot{M}=\dot{M}'$, 
respectively for an inclination angle $i = 70.2^\circ$ (bottom left panel) and $20^\circ$ 
(bottom right panel). In the case of a low inclination angle, $i = 20^\circ$, one finds
more or less the same constraints, see the top panel and the bottom right panel in 
Fig.~\ref{fig1b}. For $i = 70.2^\circ$, the difference is larger, but only for 
$\epsilon_3 > 10$.

\begin{figure}
\begin{center}
\includegraphics[height=8cm]{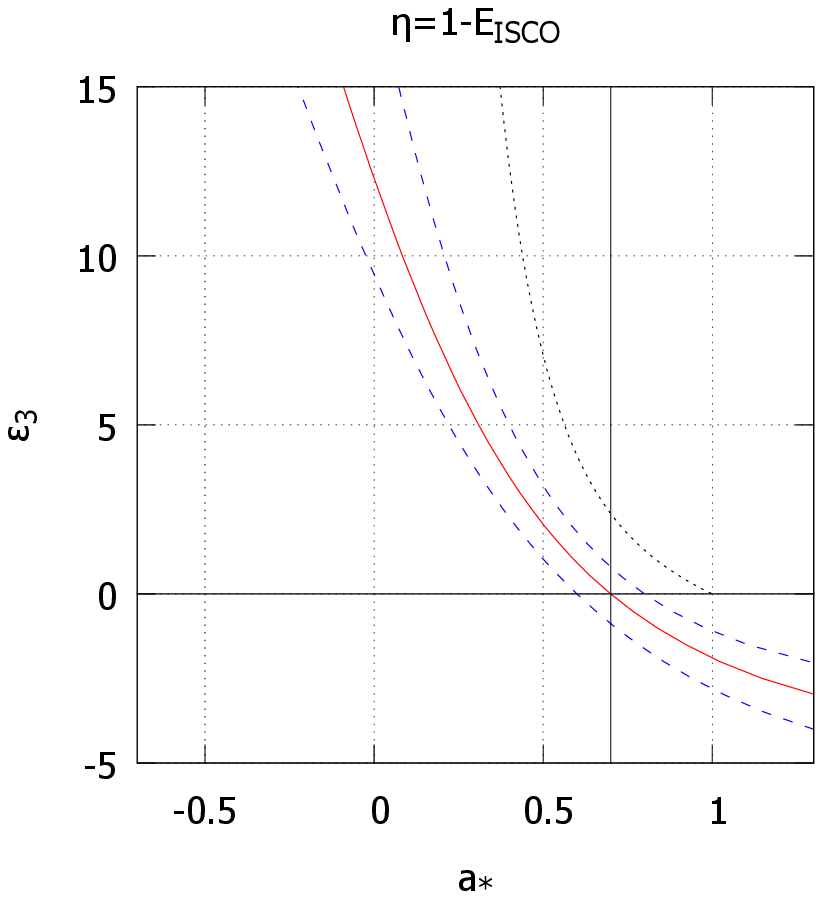}
\includegraphics[height=8cm]{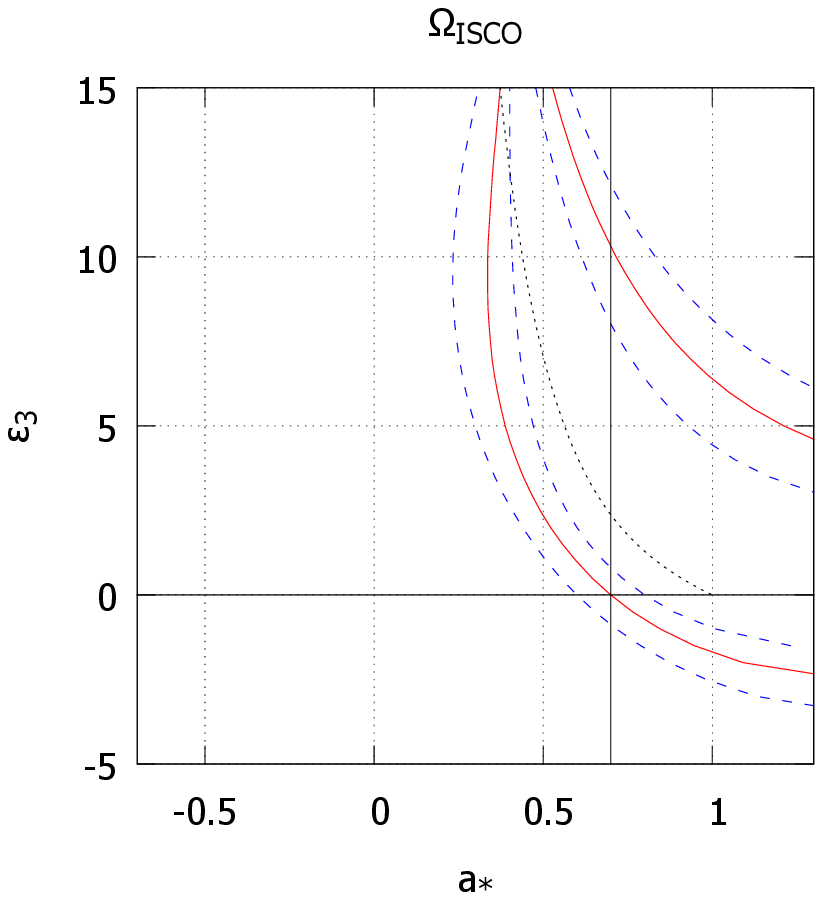}
\end{center}
\caption{Left panel: radiative efficiency in the Novikov-Thorne model 
$\eta = 1 - E_{\rm ISCO}$. The red-solid line is for $\eta = 0.1036$, which is 
the same as the one for a Kerr BH with $a_* = 0.7$. The blue-dashed
lines are for $\eta = 0.0912$ and 0.1221, which correspond, respectively,
to the Novikov-Thorne radiative efficiency in Kerr background with $a_* = 0.6$ 
and 0.8. Right panel: as in the left panel for the frequency at the ISCO radius.
The red-solid lines are for $\Omega_{\rm ISCO}/M = 0.1439$, while the 
blue-dashed lines are for $\Omega_{\rm ISCO}/M = 0.1236$ (external lines)
and 0.1737 (internal lines). In the Kerr background, these values of 
$\Omega_{\rm ISCO}/M$ correspond, respectively, to $a_* = 0.7$, 0.6, and 0.8.
See the text for more details. \label{fig1c}}
\end{figure}

Lastly, let us note that in previous work on tests of the Kerr nature of 
BH candidates we have used as crude approximation of the continuum-fitting 
measurements the values of the Novikov-Thorne radiative efficiency 
$\eta = 1 - E_{\rm ISCO}$~\citep{jet1,jet2,qpo1}; that is, at first approximation the 
continuum-fitting method actually measures the radiative efficiency in the 
Novikov-Thorne model. We thus exploited this fact translating the spin measurements 
in the Kerr background into radiative efficiency measurements, and then the latter
into spin measurements for Johannsen-Psaltis BHs with non-vanishing deformation
parameter $\epsilon_3$. In the left panel of Fig.~\ref{fig1c}, we show the constraints 
that can be obtained with this approximation for the BH candidate in GRO~J1655-40.
In the Kerr background, the spin measurement is $a_* = 0.7$ and the 1-sigma lower
and upper bounds are, respectively, $a_* = 0.6$ and 0.8. These three values of the
spin parameter correspond to $\eta = 0.1036$, 0.0912, and 0.1221 in the Kerr metric.
In the left panel in Fig.~\ref{fig1c}, the red-solid line marks the objects with 
Novikov-Thorne radiative efficiency $\eta = 0.1036$, while the blue-dashed lines
are for spacetimes with $\eta = 0.0912$ and 0.1221. The constraints obtained with
this simplification (that clearly does not take the inclination angle into account)
seem to provide a quite good result if the inclination angle of the disk is low, while
there is a more significant departure from the correct measurement in the case of 
high inclination angles. For instance, for $\epsilon_3 = 15$ one finds $a_* = -0.09 \pm 0.15$
from the $\eta$-approach, to be compared with the spin measurement 
$a_* = -0.13^{+0.10}_{-0.18}$ for the case $i = 20^\circ$ and the estimate
$a_* = 0.17\pm 0.30$ for $i = 70.2^\circ$. In the right panel in Fig.~\ref{fig1c}, we
show the contour levels for a Keplerian frequency at the ISCO radius 
$\Omega_{\rm ISCO}/M = 0.1439$ (red-solid lines) and 0.1236 and 0.1737 (blue-dashed
lines). In the case of Kerr background, these values correspond to spacetimes
with $a_* = 0.7$, 0.6 and 0.8, so the case of GRO~J1655-40. The radiative
efficiency is a better proxy for the continuum-fitting measurements.

\subsection{Slow-rotating objects: XTE~J1550-564, LMC~X-3, H1743-322, and A0620-00}

Fig.~\ref{fig2} shows the constraints for XTE~J1550-564 (top left panel), 
LMC~X-3 (top right panel), H1743-322 (bottom left panel), and A0620-00 (bottom right
panel). Even if the measurement of $a_*$ depends on the value of $\epsilon_3$, in
any case all these objects can be classified as slow-rotating. Here 
$\chi^2_{\rm min}(\epsilon_3)$ is always very close to zero and the difference
between the constraints given by the blue-dashed curves and the green-dashed-dotted
curves is definitively small. For these objects, there is a true degeneracy between $a_*$ and 
$\epsilon_3$, which is the best situation to employ our approach.

\subsection{Fast-rotating objects: M33~X-7 and LMC~X-1}

Fig.~\ref{fig2b} reports the constraints for the BH candidates in
M33~X-7 (left panel) and LMC~X-1 (right panel). These objects can be roughly
classified as fast-rotating and their constraints present some new features for large
values of the deformation parameter. For
small values of $\epsilon_3$, say $\epsilon_3 \lesssim 10$ for M33~X-7 and 
$\epsilon_3 \lesssim 5$ for LMC~X-1, $\chi^2_{\rm min}(\epsilon_3)$ is still close 
to zero and the difference between the constraints provided by the blue-dashed 
and green-dashed-dotted curves is small. For larger values of $\epsilon_3$, 
$\chi^2_{\rm min}(\epsilon_3)$ can significantly differ from zero, the bounds provided
by the blue-dashed and green-dashed-dotted curves tend to disagree, 
and actually it is possible to get a bound on $\epsilon_3$. These
constraints can be understood in terms of the location of the ISCO radius or,
more precisely, in terms of the radiative efficiency in the Novikov-Thorne model, 
namely $\eta = 1 - E_{\rm ISCO}$. The radial coordinate of the ISCO radius 
has indeed no physical meaning in general relativity, being the choice of the
coordinate system arbitrary. The continuum-fitting method eventually measure
something similar to $\eta$ and, as shown in Fig.~\ref{fig-eta}, the radiative efficiency
can be very high only for small values $|\epsilon_3|$. 
This explains the two local minima of $\chi^2$ found in the plot of M33~X-7
for $7 < \epsilon_3 < 13$ in the range of our
plane. The actual constrain for this object is given by the region between the two
blue-dashed curves. Roughly speaking, the disk thermal spectrum of the BHs inside 
the internal blue-dashed line looks more like the one of Kerr BHs with spin parameter 
larger than the upper bound inferred assuming the Kerr background, namely
$a_* = 0.89$. The spectrum of the BHs outside the external blue-dashed line is more 
similar to the one of Kerr BHs with spin parameter lower than the lower bound inferred 
assuming the Kerr metric, $a_* = 0.79$. If we assume the Kerr background and we 
analyze the X-ray data of the thermal spectrum of a thin disk in a Johannsen-Psaltis 
background located outside the exterior blue-dashed curve, we would obtain a lower 
spin parameter than the one measured. The 1-sigma bound $a_* > 0.79$ obtained 
in the Kerr background thus implies a spin independent constraint $\epsilon_3 < 14$. 
While the red-solid line for $\epsilon_3 > 14$ above the connection point is still
the minimum of $\chi^2$, the spectrum of these objects are more similar to the ones
of Kerr BH with spin $a_* < 0.79$.

The constraints for LMC~X-1 are actually similar, even if they look 
to be very different at a first sight. The difference in the allowed regions of M33~X-7 and 
LMC~X-1 is due to the different inclination angle (high $i$ for M33~X-7, low $i$ 
for LMC~X-1), the higher value of the spin parameter inferred for LMC~X-1, and 
the precision of the Kerr measurement. Because of the lower inclination angle, 
LMC~X-1 has not two local minima of $\chi^2$ in the range under consideration 
(we would have found the second minimum, as well as the connection point 
between the two minima, if we had included higher values of $\epsilon_3$). 
Moreover, the blue-dashed curve closes at $\epsilon_3 \approx 11$ because 
the spectra in spacetimes with higher deformation parameter have $\chi^2$ 
larger than $\sigma_-$ and $\sigma_+$\footnote{We note that
the blue-dashed curve closes: it may not be clear from the picture simply
because it overlaps with the red line.  } 
If the Kerr spin measurement of 
LMC~X-1 had a larger uncertainty than the 1-sigma bound $0.85 < a_* < 0.97$,
we would have obtained a larger allowed region on the spin parameter--deformation
parameter plane, the blue-dashed curve would have not closed, and the 
constraint would have looked more similar to the one of M33~X-7.  
If, for instance, the BH candidate 
in LMC~X-1 were a Johannsen-Psaltis BH with $\epsilon_3 = 13$, the analysis 
of the disk's thermal spectrum under the assumption of Kerr background would 
provide the spin parameter reported in Tab.~\ref{tab} for the objects on the red-solid 
line, but the fit would not be good. In the case of LMC~X-1, the 
analysis of real data may distinguish a Kerr BH from one with $\epsilon_3 > 11.5$.

\subsection{Very fast-rotating objects: GRS1915+105 and Cygnus~X-1}

\begin{figure}
\begin{center}
\includegraphics[height=8cm]{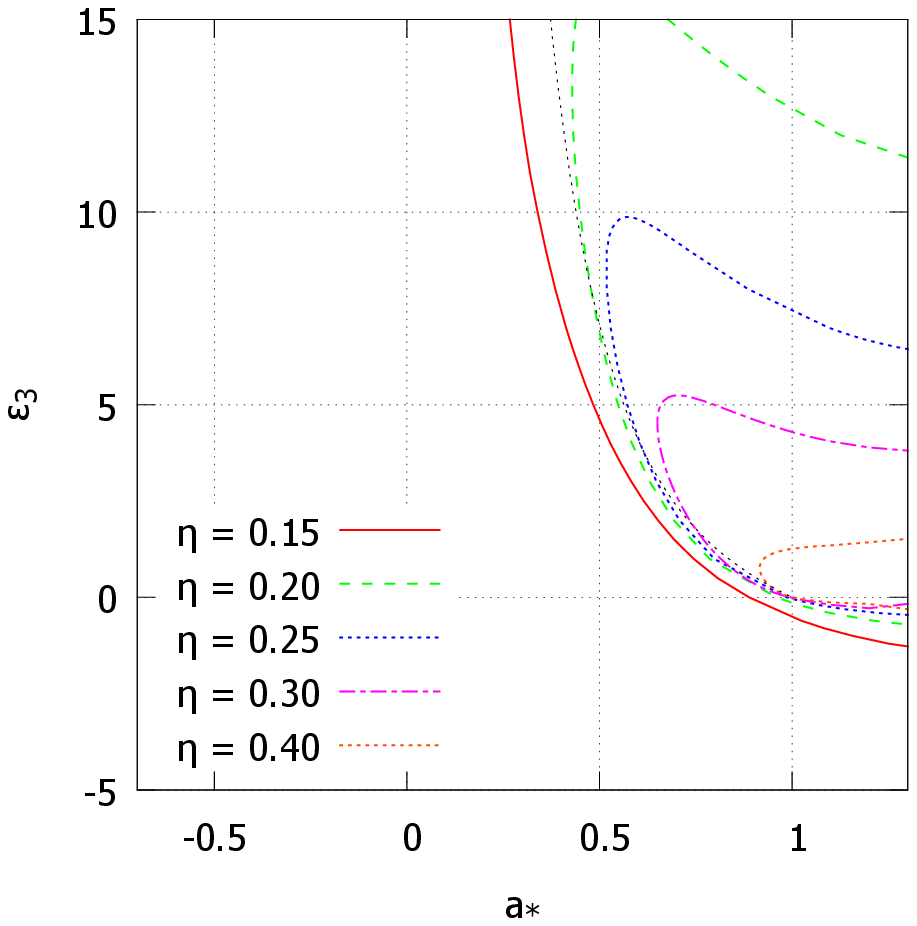}
\end{center}
\caption{Some contour-levels showing the radiative efficiency in 
the Novikov-Thorne model $\eta = 1 - E_{\rm ISCO}$. 
This is roughly the actual physical quantity measured in the 
continuum-fitting method, which can be later translated into a spin measurement 
in the Kerr background because there is a one-to-one correspondence between 
$\eta$ and $a_*$. In the case of non-Kerr background, one finds an allowed 
region on the spin parameter--deformation parameter plane, since 
$\eta = \eta (a_*, \epsilon_3)$. Such an approximation works quite well for small 
viewing angles and low $\eta$, while there are some quantitative differences for 
large viewing angles and high $\eta$. See the text and Footnote~2 for more details. 
\label{fig-eta}}
\end{figure}

For the BH candidates in GRS1915+105 and in Cygnus~X-1, the Kerr
spin measurement is just a lower bound, $a_* > 0.98$. The constraints on the spin
parameter--deformation parameter plane are reported in Fig.~\ref{fig3} and they have
been obtained following a slightly different approach, which is dictated by the fact
that we have a lower bound instead of a measurement. The plots only show the
blue-dashed lines, which here separate BHs whose spectrum looks more similar to 
the one of a Kerr BH with spin parameter higher/lower than 0.98. More precisely, 
the allowed regions are the ones inside the blue-dashed lines (the spectra of these 
BHs look like the one of a Kerr BH with spin parameter larger than 0.98), while the 
regions outside are ruled out (the spectra of these BHs would be interpreted as the 
one of a Kerr BH with spin lower than 0.98). The difference between the constraints
of GRS1915+105 and Cygnus~X-1 is only due to their different viewing angle.
The remarkable difference with respect to the other BH candidate constraints is that 
here the deformation parameter $\epsilon_3$ turns out to be well constrained. This 
happens because Johannsen-Psaltis BHs can mimic very fast-rotating Kerr BHs 
for a restricted range of the spin and the deformation parameters [but this is not true
for other kinds of deformations, see~\citet{v2-cb}]. In the case of GRS1915+105, the 
Kerr bound $a_* > 0.98$ implies $\epsilon_3 < 6.0$. For Cygnus~X-1, the constraint 
is $\epsilon_3 < 4.0$. Again, these constraints could be expected on the basis of 
the radiative efficiency of the Johannsen-Psaltis spacetimes shown in 
Fig.~\ref{fig-eta}\footnote{For high values of $a_*$ and $\eta$, there is some discrepancy 
at the quantitative level between the constraints imposed by the two parameters. For 
instance, the Kerr spin estimate $a_* > 0.98$ would require $\eta > 0.234$, while 
from the comparison of Fig.~\ref{fig3} and Fig.~\ref{fig-eta} it seems that the 
measurement $a_* > 0.98$ is more similar to a bound such as $\eta > 0.30$.
The constraints would have been very similar in the case we had imposed
$\dot{M} = {\rm constant}$, 
as it can be seen from the comparison of the right panel in 
Fig.~4 in~\citet{rev2} (where $\dot{M} = {\rm constant}$) and Fig.~\ref{fig-eta} in the
present paper. Here the analysis has been conducted assuming
$\eta \dot{M} = {\rm constant}$ for the reasons discussed in Section~\ref{s-con}.}.
We note, however, that the allowed regions for GRS1915+105 and Cygnus~X-1
reported in Fig.~\ref{fig3} have also some very pathological spacetimes in which there
is no ISCO and the radiative efficiency can be very high (potentially mimicking Kerr 
metrics with $a_* > 1$, but current measurements cannot rule out the possibility that
these sources have a radiative efficiency exceeding the one of a Kerr BH with $a_* = 1$, 
see the original papers).


\section{Summary and conclusions} \label{s-sc}

Astrophysical BH candidates are thought to be the Kerr BHs predicted in general 
relativity, but there is not yet an observational confirmation that the spacetime 
geometry around them is described by the Kerr solution. The observation of 
features associated to relativistic effects in their spectra is not enough to confirm 
the Kerr nature of these objects, because non-Kerr BHs may look like Kerr BHs 
with different spin~\citep{cfm-iron,shadow1,spot,shadow2}. To test the Kerr BH 
hypothesis, we have to consider a more general background that includes the 
Kerr solution as special case, and check whether observational data only
require the Kerr solution.

The continuum-fitting methods is the analysis of the thermal spectrum of geometrically 
thin and optically thick accretion disks. Assuming that the spacetime around BH 
candidates is described by the Kerr metric, this technique is used to infer the BH 
spin parameter $a_*$. In the present paper, we have ``translated'' the spin 
measurements reported in the literature into constraints on the spin parameter--deformation
parameter plane. This has been achieved by comparing Kerr models with the spin 
parameters reported in the literature with the spectra computed in the Johannsen-Psaltis 
metric. The approach is possible thanks to a fundamental degeneracy between
the spin and possible deviations from the Kerr solution. This significantly simplifies
the analysis, since we can just focus our attention on the role of the spacetime metric, 
assuming that all the astrophysical effects and the instrumental issues have been 
already properly taken into account in the previous studies.

The results of this work are Figs.~\ref{fig1}-\ref{fig3}. The constraints obtained here 
will be combined with other measurements to break the degeneracy between the
spin parameter and the deformation parameter. The iron line profile is currently
the only other relatively robust technique to probe the spacetime geometry around 
BH candidates and it has been already extended to non-Kerr 
backgrounds~\citep{iron-jp,iron-cb}. However, only for a few sources we have good 
X-ray data for both the continuum-fitting method and the iron line profile. 
It is also to be noted that both the techniques strongly rely on the fact that the 
inner edge of the disk is at the ISCO radius, so their combination may not be the 
best choice to break the degeneracy and only very good measurement -- not 
available today -- could put interesting constraints~\citep{cfm-iron}. A more 
promising approach could be the combination of the continuum-fitting constraints 
with current bounds from quasi-periodic oscillations (or QPO)~\citep{qpo-jp,qpo1,qpo2}, 
but at present it is not clear the exact mechanism responsible for these phenomena 
and different models predict different measurements, which makes this 
technique not yet mature to test fundamental physics.


\begin{acknowledgments}
We thank Lijun Gou and Jack Steiner for useful
discussions and suggestions. 
This work was supported by the NSFC grant No.~11305038, the 
Shanghai Municipal Education Commission grant for Innovative 
Programs No.~14ZZ001, the Thousand Young Talents Program, 
and Fudan University.
\end{acknowledgments}


\end{document}